\documentclass[a4paper]{article}
\usepackage{graphicx}

\begin{document}

\title{TRTViewer: the ATLAS TRT detector monitoring and diagnostics tool}
\author{S.Yu.~Smirnov\footnote{On behalf of the ATLAS Collaboration}\\
  National Research Nuclear University "MEPhI",\\
  115409 Moscow, Russia\\
  \texttt{Serge.Smirnov@cern.ch}}

\date{}
\maketitle

\begin{abstract}
The transition radiation tracker (TRT) is the outermost of the three sub-systems of the ATLAS inner detector at the Large
Hadron Collider (LHC) at CERN.
It is designed to combine the drift tube tracker with the transition radiation detector, providing an important contribution to the
charged particles precise momentum measurement and particle (mainly electron) identification.
The TRT consists of a barrel section at small pseudorapidity ($\eta$) and two separate end-cap partitions at large $\eta$.
The detector performance and its operational conditions were permanently monitored during all commissioning and data-taking
stages using various software tools, one of which -- TRTViewer -- is described in the present paper.
The TRTViewer is the dedicated program for monitoring the TRT raw data quality and detector performance at different hardware
levels: individual straws, readout chips and electronic boards.
The data analysis results can be presented on the event-by-event basis or in the form of color maps representing the
operation parameters (efficiencies, timing, occupancy, etc.) according to the real geometrical position of the detector hardware
elements.
The paper describes the TRTViewer software package as the event displaying tool, raw data processor and histogram and
operation parameters presenter, which works with the different sources of input information: raw data files, online monitoring
histograms, offline analysis histograms and TRT DAQ Configuration database.
The package proved to be one of the main instruments for the fast and effective TRT diagnostics during debugging and operation
periods.
\end{abstract}

\section{Introduction}

The transition radiation tracker (TRT) \cite{TRTbarrel08, TRTendcap08} is the combined gaseous detector occupying the
outermost part of the inner detector of the ATLAS \cite{ATLAS08} -- one of the two general purpose experiments at the Large
Hadron Collider at CERN.
The inner detector is located in the center of the ATLAS experiment, surrounding the beam interaction point.
It is immersed in a 2 T magnetic field generated by the central solenoid.
The TRT has a cylindrical shape 2.1 m in diameter and 6.2 m long.
It consists of close to 300,000 thin-wall drift tubes (straws) \cite{TRTstraw08}, providing on average 30 two-dimensional
space points for charged particle tracks.
The straws are 4 mm in diameter and their length varies from 39 cm in end-cap partitions to 144 cm in the barrel section.
The main TRT task is to provide accurate charge particle track measurement with 0.12-0.15 mm resolution, significantly
contributing to the tracking performance and particle momentum reconstruction of the inner detector as a whole,
particularly at high transverse momentum $p_T$.
The particle momentum is measured for tracks with pseudorapidity $\mid\eta\mid < 2.0$ and $p_T > 0.5$ GeV.
Along with continuous tracking, the detector ensures particle identification capability through the detection of transition
radiation X-ray photons generated by high velocity particles in many polymer fibers or films that fill the spaces between
the straws.
The latter functionality provides substantial discriminating power between electrons and hadrons in the momentum range
from 1 to 200 GeV.
More detailed TRT description and recent performance results can be found in the other contributions to the present
Conference proceedings \cite{Adelman11, Hines11}.

The TRT was fully installed in its final position in ATLAS in August, 2008.
However the detector commissioning had already begun during the first assembly stages \cite{Degenhardt10}.
These early commissioning stages consisted mainly of checking out electronics, threshold tuning and determination of the
time delays.
The first signals and reconstructed tracks from cosmic particles were already observed in the TRT in 2005.
During the first injections of protons into the LHC in 2008 and 2009, beam splashes were used to check the timing alignment
of the TRT.
Millions of cosmic ray events were collected with the specially developed TRT Fast-OR trigger \cite{Fratina09}, which were
essential in aligning and timing-in the whole ATLAS detector.
Then during the LHC commissioning in the autumn of 2009, the good timing properties of the TRT were used to confirm
the first proton-proton collisions in the ATLAS detector.
Finally, the initial collision runs were used to continue to improve the alignment, as well as to calibrate the response of the TRT
to transition radiation photons from electrons and charged hadrons.
The TRT has shown excellent performance remaining operational for 100\% of the 2009, 2010 and 2011 physics runs,
having a better than 94\% hit efficiency, good discrimination between charged hadrons and electrons as well as
providing particle discrimination based on ionization loses.
The latest alignment and calibration studies \cite{Alignment11, Calibration11} show that the spatial hit resolution of the
TRT is well consistent with the design specification and it is 118 $\mu$m (barrel) and 132 $\mu$m (end-caps) for the
high transverse momentum tracks.
This leads to a significant improvement of the inner detector momentum resolution by the TRT.

During all the commissioning and data-taking stages, the TRT performance and hardware operational conditions were
continuously monitored using various software tools.
One of them -- TRTViewer is a dedicated program for monitoring the TRT raw data quality and detector performance
at different hardware levels: individual straws, readout chips, receiving signals from 16 straws, and electronic boards,
on which from 10 to 27 chips are placed depending on the board type.

\section{TRTViewer description}

The running of the large scale detector such as the TRT requires the constant monitoring of many operational parameters
in order to ensure the high quality of the delivered physical data and the detector safety.
The TRT team has developed the monitoring and analysis software package -- TRTViewer -- as a convenient and
flexible tool for fast and effective detector diagnostics and debugging.
The main TRT software design concept is to combine in one application: the event displaying tool, the raw data processor
and the histogram and operational parameters presenter.
The presenter displays graphical information according to the real geometrical positions of TRT different hardware
components.
TRTViewer runs under the CERN baseline operational system Scientific Linux CERN v.5 \cite{linux} and uses ROOT
graphical libraries \cite{root} which are widely used in the high energy physics community.

The application operates in four different modes depending on the source of the input information: raw data (files and
direct stream from Data Acquisition System -- DAQ), online monitoring histograms (via ATLAS Online
Histograming/Information Service), offline analysis histograms (Monitoring Data Archive) and TRT DAQ Configuration
Data Base.
It uses a simple text configuration file for setting parameters and choosing between the running modes.

The TRTViewer can display all three detector partitions: barrel and two end-caps.
The barrel section is shown from both sides because the straws in this part are being read out from both ends
independently.
The image of the barrel represents the plane, perpendicular to the beam axis, giving the $R - \phi$ coordinates of straws.
The end-cap parts consist of the straws pointing radially outwards from the beam axis, thus producing the
$Z - \phi$ coordinates of charged particle hits.
The end-caps are shown as two unfolded surfaces corresponding to the outer radius of end-cap cylinder, where the
read-out electronics is located.

\begin{figure}[tbh]\begin{center}
  \includegraphics[height=14cm]{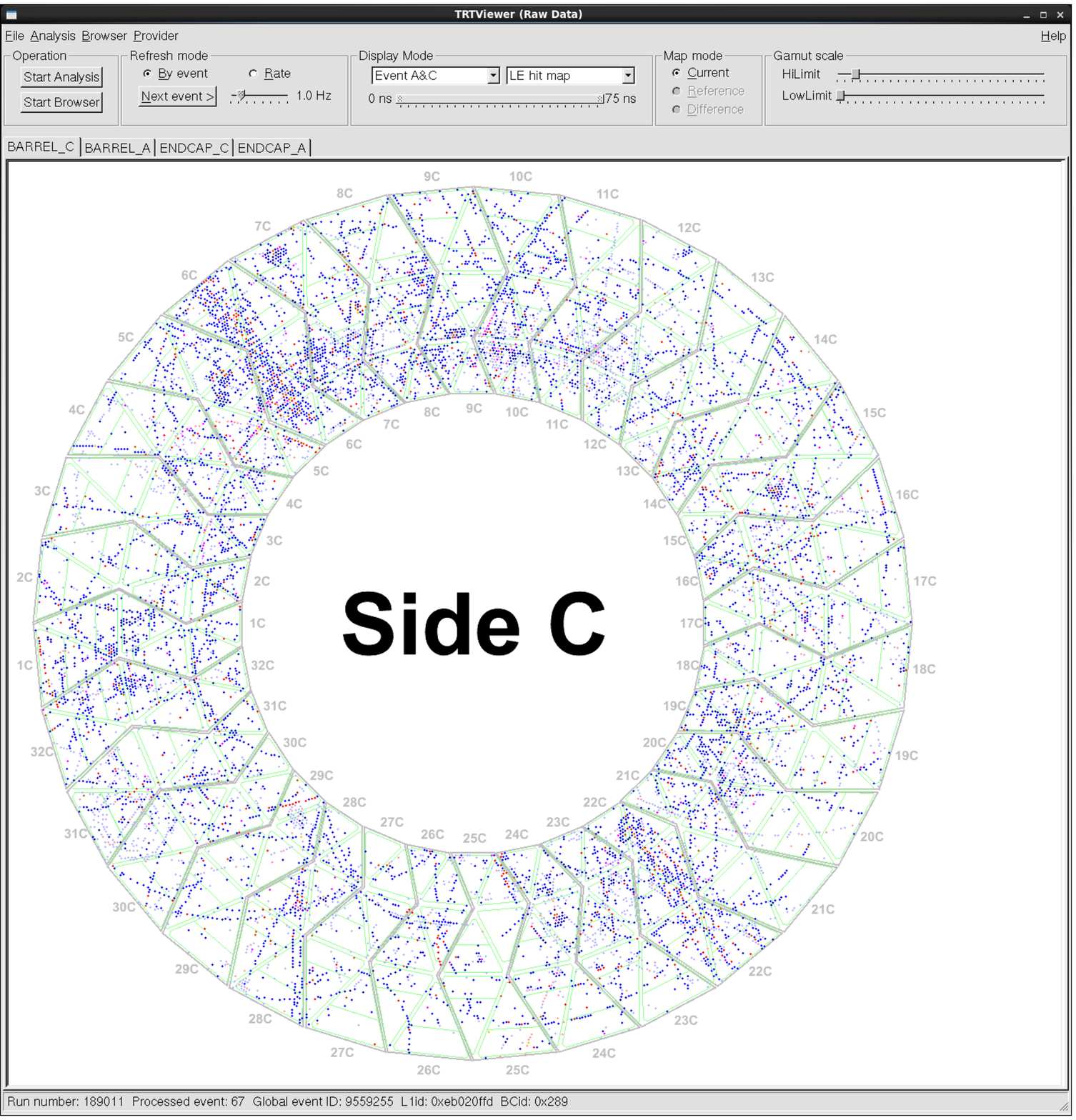}
  \caption{The main window and graphical user interface of TRTViewer in Raw Data mode.}
  \label{fig1}
\end{center}\end{figure}

The graphical interface of the application has the following visualization modes.
First an event display working at the level of the smallest detector components (straws).
And second a histogram and color map presenter operating with electronics boards, readout chips and straws.
When processing the raw data, it is possible to display full information from the current event, including zooming to
the individual drift tubes and showing the straw signal time evolution diagram over 75 ns, which corresponds to three
consecutive LHC bunch crossings.
The straw signal pulse height is coded by color: blue dots correspond to the signal above the low threshold
($>$300 eV), which is used for detecting the passage of minimum ionizing particles, and the red ones correspond to the high
threshold ($>$6 keV), which flags the absorption of transition radiation X-rays in a straw. 
The TRTViewer track fitting algorithm can be configured to fit both straight tracks when the magnetic field is off
or curved ones when the field is on.
In both cases the tracks can be drawn on top of the corresponding event (fig. \ref{fig1}).

\begin{figure}[tbh]\begin{center}
  \includegraphics[height=14cm]{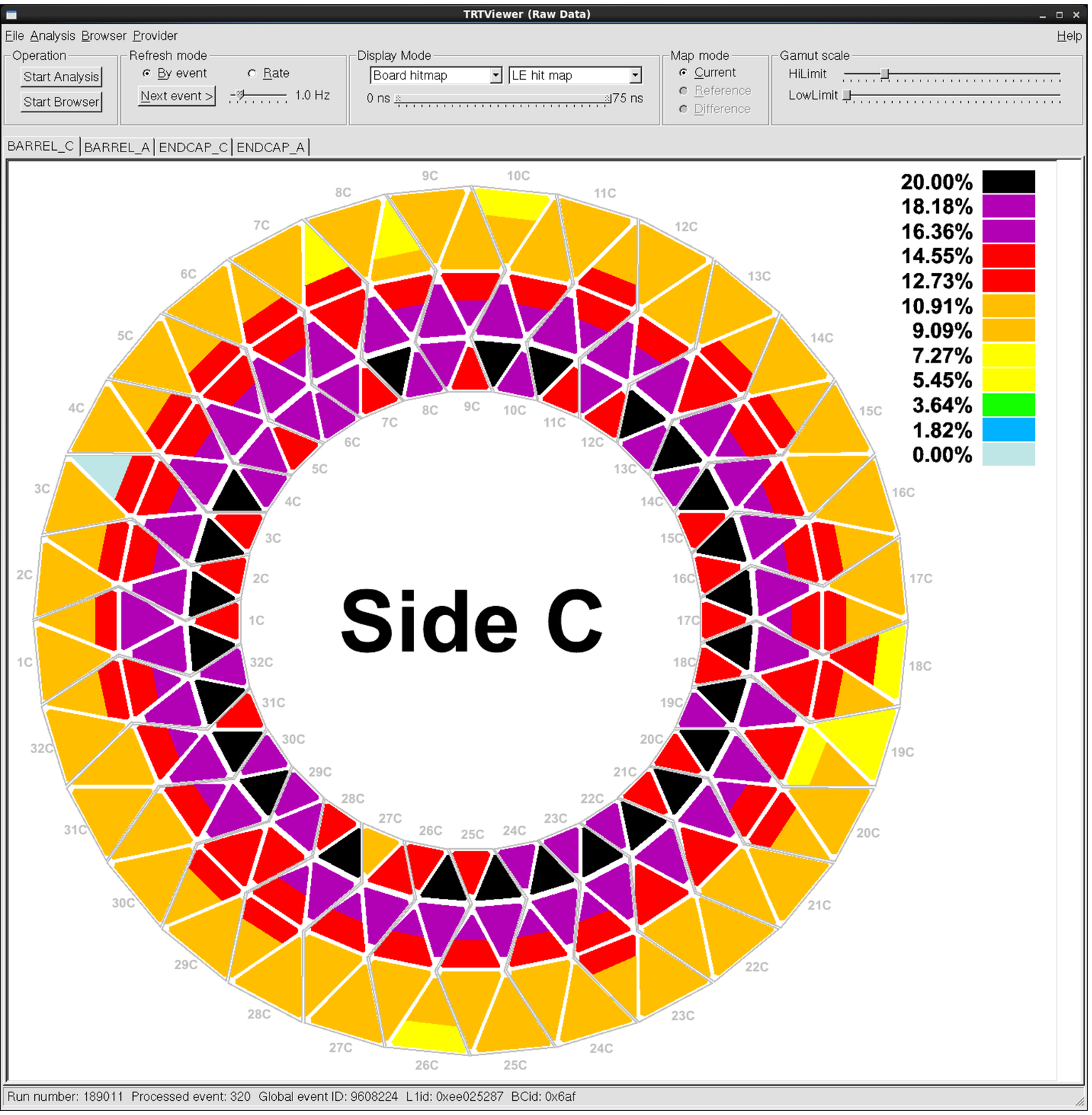}
  \caption{An example of electronics boards color map showing the average board occupancy for low threshold hits.}
  \label{fig2}
\end{center}\end{figure}

The raw data is processed by the physics analysis module and the results are stored in the form of ROOT histograms.
They can be immediately presented by the standard ROOT browser in the form of ordinary histograms or can be shown
in the form of color maps, where an operational parameter or a physical value (e.g. efficiency, timing, occupancy, etc.)
is coded by the color, according to an interactively chosen scale, and the image of the corresponding hardware
element (straw, readout chip or electronics board) is drawn using this color at the real geometrical position of the detector
element (fig. \ref{fig2}).
The same color map technique is used for the other modes of TRTViewer operation: presentation of online monitoring
histograms and offline analysis histogram archives.
The only difference is that the analysis histograms are not produced by the TRTViewer but instead are read from
ROOT files which have been created by other TRT monitoring applications.
These monitoring applications run under the ATLAS general software framework ATHENA \cite{Athena} and use different
analysis algorithms but have a common output histogram format and ROOT subdirectory structure.

The TRT DAQ Configuration database is used to store the readout electronics and trigger settings that change fairly often
such as fine timing delays and signal amplitude thresholds.
These values are taken from the database by the corresponding TRTViewer module and they are reformatted in order
to be consistent with the format required by the color maps technique.
The database module gives also the possibility to interactively edit and write back to the database the settings of the
detector read out electronics, thus ensuring the fast and convenient control over the detector performance.

 The TRTViewer is supplied with powerful analytical tools which proved to be very helpful to the detector experts during
 the commissioning and data-taking stages.
 Some of them are implemented as pure graphical interface tools and others are supported by the physics analysis
 algorithms built-in into the application.
 For example, double clicking on the detector image brings up the zoomed view, which can be further enlarged or
 shifted to another detector region.
 This is very helpful for studying particular faulty TRT straws.
 One can easily figure out which readout element a faulty straw is connected to and its influence on the performance of
 the adjacent straws and chips, etc. 
One can also display the hit arrival time distribution for a straw by left clicking with on that straw in the zoomed view of straw
color map.
 To study signal overlapping (shadowing) from adjacent LHC bunch crossings, a special time window was 
 introduced which skips analysis of straw hits outside time window.
 The window is controlled by a double slider, giving the possibility to change interactively its
 position in time and duration within three bunch crossings.
 
 The most useful TRTViewer analysis of TRT performance is to study its development over time.
 The TRTViewer has the possibility to choose and load any ROOT histogram file, produced by any other TRT
 monitoring application, and use it as a reference file for comparison with the current data.
 For histogram comparison one can choose several presentation modes: current histogram only, current
 histogram overlapped with the reference one, difference between current and reference histogram and their ratio.
 In the color maps comparison mode it is also possible to display the difference or the ratio between
 the current and reference maps.
 
\section{Conclusion}

The ATLAS TRT team has developed an efficient and multifunctional software tool TRTViewer.
It has proved to be one of the most helpful instruments for the fast and effective TRT diagnostics during debugging,
commissioning and operational stages.
The TRTViewer serves simultaneously as event displaying tool, raw data processor with simple reconstruction and
physics analysis algorithms, histogram and operational parameters presenter.
The application works with different sources of input data: raw data files, direct DAQ data streams, online monitoring
histograms, offline analysis histogram archives and the TRT DAQ Configuration database.
The data analysis results can be shown on the event-by-event basic or in the form of color maps where the
operational parameters (efficiencies, occupancies, maps of noisy or dead detector elements) are represented according
to the real geometrical position of the detector parts: individual straws, readout chips and electronics boards.

\end{document}